\documentclass[12pt]{article}
\usepackage{latexsym,amsmath, graphicx}
\begin{document}
\begin{center}

{\bf REPLY TO "COMMENT ON 'ANALYTICAL AND NUMERICAL VERIFICATION
OF THE NERNST THEOREM FOR METALS' " }

\vspace{1cm}
 Johan S. H{\o}ye\footnote{E-mail:
johan.hoye@phys.ntnu.no}

\bigskip

Department of Physics, Norwegian University of Science and
Technology, N-7491 Trondheim, Norway

\bigskip

Iver Brevik\footnote{E-mail: iver.h.brevik@ntnu.no} and Simen A.
Ellingsen\footnote{E-mail: simen.a.ellingsen@ntnu.no}

\bigskip

Department of Energy and Process Engineering, Norwegian University
of Science and Technology, N-7491 Trondheim, Norway

\bigskip

Jan B. Aarseth\footnote{E-mail: jan.b.aarseth@ntnu.no}

\bigskip

Department of Structural Engineering, Norwegian University of
Science and Technology, N-7491 Trondheim, Norway

\bigskip

\end{center}

\bigskip

\begin{abstract}

In this Reply to the preceding Comment of Klimchitskaya and
Mostepanenko (cf. quant-ph/0703214), we summarize and maintain our
position that the Drude dispersion relation when inserted in the
Lifshitz formula gives a thermodynamically satisfactory
description of the Casimir force, also in the limiting case when
the relaxation frequency goes to zero (perfect crystals).

\end{abstract}


\newpage
The main concern of the authors of Ref.~\cite{klimchitskaya07} is
that in their opinion the Drude dielectric function is not valid
if the metal is a perfect crystal, with no impurities at all. By
contrast, if the metal is {\it imperfect} implying some
relaxation, the authors seem to agree with our standpoint as
expressed in Ref.¨\cite{hoye07}: The Drude dispersion relation is
adequate to describe the medium and there is no conflict with the
Nernst theorem in thermodynamics. The Drude relation implies that
there is a finite relaxation frequency, which we call $\nu$. We
find it satisfactory that now there seems to be full agreement as
far as imperfect crystals are concerned. The problem remains for
perfect crystals only.

We can agree with the claim of Ref.~\cite{klimchitskaya07} that
perfect crystals is an idealization that forms much of the basis
of solid state physics. From a physical viewpoint taking the limit
$\nu \rightarrow 0$ is however in the Casimir context a delicate
and not unique operation. There are several ways to argue that our
results found via use of the Drude relation are valid in general.

\begin{itemize}
\item[1]  In practice,  all solids have some relaxation. This
follows also from the Kramers-Kronig relations, for any material
having frequency dispersion. Ideal crystals can be seen as the
limit of no relaxation $\nu\rightarrow0$. We have shown very
accurately, both analytically and numerically \cite{hoye07} - cf.
also Refs.¨\cite{brevik07,hoye03} - that for $\nu>0$ the entropy
$S=0$ at $T=0$. From this we draw the conclusion that in the limit
$\nu\rightarrow 0$ the value $S=0$ is preserved. But  the entropy
decreases more and more rapidly when $T$ increases from $T=0$, and
this decrease becomes a discontinuous jump in the limit
 $\nu\rightarrow 0$. The variation of the entropy as a function of $T$ for two
 copper half-spaces is nicely
 illustrated in Fig.~3 in the QFEXT03 Proceedings article of
 Sernelius and Bostr\"{o}m \cite{sernelius04}, for a finite value of $\nu$.

\item[2] Further, merely taking the limit $\nu\rightarrow 0$ is not
fully satisfactory from a physical viewpoint. It should be noted that
on the molecular level the
permittivity becomes non-local,
$\varepsilon=\varepsilon(\omega, k)$, where $k$ is the magnitude
of the wave vector $\bf k$. Then $\varepsilon$ becomes finite for
finite $\bf k$. Only the special case $\varepsilon(0,0)$ is
infinite, and this ("measure zero") can not be expected to
give a finite contribution to the Casimir
force.
In independent studies Svetovoy and Esquivel \cite{svetovoy05} and Sernelius \cite{sernelius05} find that
 the TE zero mode should not contribute when spatial dispersion is carefully taken into account. Because
 of nonlocal effects, the former of these state, ``the question does [$\nu$] go to zero or have some residual
  value at $T\to 0$ becomes unimportant'' and they find that the TE zero frequency contribution must indeed be
   zero to satisfy Nernst's theorem.

 \item[3] A TE zero frequency mode is
not a solution of Maxwell's equations and should  not occur for
that reason. This point is discussed in more detail in
Refs.~\cite{hoye03} and \cite{hoye01}. (See especially Sec.~III of
Ref.~\cite{hoye01}.)

\item[4] Introducing a TE zero frequency mode for ideal crystals
would imply that a medium with $\nu=0$ would behave differently
from a real medium when taking the limit $\nu\rightarrow0$. In the
former case with the TE-mode included the entropy increases slowly
from $S=0$ when $T$ increases, while in the latter case it
decreases rapidly to negative values for a real medium with
$\nu>0$. Thus in the limit $\nu\rightarrow 0$ the result of the
latter is different from the former. Due to this the use of the
latter for $\nu\rightarrow0$ and the former for $\nu=0$ would
create a discontinuity in physical properties at $\nu=0$ that we
find unphysical. \vspace{0.4cm}


In connection with Eq.\ (3) in Ref.\ \cite{klimchitskaya07} it is argued that $\nu(T)\ll \zeta_m(T)=2\pi kmT/\hbar$ $(m=1,2,3, \cdots)$ holds as temperature decreases. Surely, this is so for a perfect lattice in which relaxation follows the Bloch-Gr\"{u}neisen formula all the way to zero temperature. We consider real metals with impurities, however, in which $\nu$ reaches some residual value at low temperatures. At temperatures so low that relaxation is dominated by impurities, their equation (3) no longer holds, but for sufficiently small $T$, Eq.\ (1) does. The authors of \cite{klimchitskaya07} explain in great detail why, for this reason, our calculated low temperature correction is numerically inaccurate. This is true, but immaterial, since our analysis is phenomenological and does not aspire to provide exact numerical estimates. The numerical calculation presented was motivated by a wish to confirm the correctness of our analytical result, which it convincingly does.

The question becomes how to define the perfect lattice at zero temperature since the limits $\nu\to 0$
 and $T\to 0$ do not commute. One may put $\nu= 0$ right at the outset, or take the limit of vanishing
 relaxation frequency subsequent to calculating the free energy. For reasons discussed above we suggest
  the latter procedure is the more appropriate. If equation (3) of \cite{klimchitskaya07} is to hold at
  all temperatures, the former procedure must be employed, which results in a different prediction of the
   Casimir force and fulfillment of Nernst's theorem by inclusion of the TE zero frequency mode.

Let us sum up: The basic assumption underlying our calculation
in Ref.~\cite{hoye07} was that the relaxation frequency $\nu$
stays {\it finite} at any temperature including $T=0$. At low
temperatures we expect $\nu$ to be  smaller than at room
temperature; this changes our results quantitatively but not
qualitatively. On the basis of this assumption, we showed that
there is no thermodynamic inconsistency associated with the Drude
relation. Now, simply setting $\nu=0$ makes our formalism to some
extent indeterminate. In particular, this indetermination persists
if $\nu=0$ for $T=0$. Physical arguments, like the requirement of
continuity of behavior with respect to different values of input
parameters, suggest strongly, however, that the results obtained
from the Drude relation are quite general.

\end{itemize}

\end{document}